# Diversity and Collective Action

Bernardo A. Huberman and Natalie S. Glance

Dynamics of Computation Group
Xerox Palo Alto Research Center
Palo Alto, CA 94304

## Abstract

We elucidate the dynamics of ongoing collective action among intentional agents with diverse beliefs and imperfect information. Their decisions on whether or not to contribute to the collective good depend not only on the past but also on their expectations as to how their actions will affect those of others. We show that in attempts at collective action the onset of overall cooperation can take place in a sudden and unexpected way. Likewise, defection can appear out of nowhere in very large, previously cooperating groups. These outbreaks mark the end of long transient states in which defection or cooperation persists in groups that cannot sustain it indefinitely. Computer experiments demonstrate these predictions, as well as verifying that diversity of beliefs among individuals acts as an additional source of uncertainty, instigating the outbreaks.

# 1 Introduction

Collective action problems create difficult quandaries for societies, be they social, economic, or organizational. Whenever a group of intentional agents collaborates in a collective task that produces some overall utility to the group, the potential for a dilemma arises. In cases where the gain to an individual for collaborating is less then its cost for participating, it may rationally choose not to cooperate and instead to free ride on the effort of others. When this logic holds for all individuals, no sustained cooperation ensues. The problem occurs when the benefit to be accrued by overall cooperation offsets individual costs, since rationality on the part of each individual leads to failure in achieving a collective good beneficial to all. The resolution of such dilemmas, which are the subject of study in the social sciences [1–6], underlies successful attempts at ongoing collective action, such as the functioning of large organizations, the adoption of new technologies [7], and the mobilization of political movements [8].

In human societies, an essential element contributing to the likelihood of collective action is that individuals differ from each other both in their beliefs and in their estimates of the costs and benefits of contributing to the collective good. Thus, whereas one individual might decide to participate in the group effort on the basis of a few others having joined, another might wait until many others have already done so. This diversity comes into play in many instances of collective action.

In studying collective action one may then ask the following question: if agents make decisions on whether or not to cooperate on the basis of imperfect information about the group activity, and incorporate expectations on how their decision will affect other agents, can overall cooperation be sustained for long periods of time? Moreover, how do expectations, diversity of beliefs, and group size affect cooperation?

Collective action problems are characterized by the impossibility of exclusion; that is, no member of a group engaged in collective action can be excluded from enjoying the benefits of the group's efforts. On the other hand, collective goods can have varying amounts of jointness of supply, which is the degree to which one agent's consumption of the good does not reduce the amount available to any other. The amount of jointness determines in part the dependence of cooperative outcomes on the size of the group [5]. Thus, for example, some studies have shown that overall cooperation is undermined as the group increases in size [1, 9], others that, to the contrary, that it is more likely for larger groups [8]. The latter result was obtained for public goods, a special subset exhibiting perfect jointness of supply, while the former holds for divisible goods, those with limited jointness of supply. In this paper, we consider divisible collective goods.

Regardless of whether or not a group of a given size can exhibit ongoing cooperation, there remains the issue of how is it that such a state is reached, if ever. For one can imagine situations whereby a group initially exhibits cooperative behavior in spite of its being too large to do so, only to gradually evolve into collective defection. Conversely, a large non-cooperative group could undergo a drastic reduction in size and the relevant question then becomes how long before the switch to global cooperation occurs.

In order to answer these questions, we present and study a dynamical model of ongoing collective action among intentional agents whose choices depend not only on the past but also on their expectations as to how their actions will affect those of others. In this model agents act on the basis of information that can be uncertain at times. We show that under these conditions the onset of overall cooperation can take place in a sudden and unexpected way. Likewise, defection can appear out of nowhere in very large, previously cooperating groups. These outbreaks mark the end of long transient states in which defection or cooperation persists in groups that cannot sustain it indefinitely. Moreover, we study the effects of diversity on these phenomena. First of all, we find that diversity acts as an additional source of uncertainty, thus shortening the time to an outbreak. Also, we show that when several subgroups with different beliefs are merged into a larger group, overall cooperation (or defection) can appear in stages.

Following a long tradition [10, 3, 9], in Section 2 we study the free rider problem as a repeated *n*-person prisoners' dilemma, wherein the benefit obtained by an individual from cooperating in producing the good is outweighed by the cost of cooperation for the one-shot game. We recast the interaction as an asynchronous dynamic game in which each individual reconsiders its decision at an average reevaluation rate $\alpha$, using delayed information of the level of provision of the good. The iterated game is of finite duration, and individuals decide to cooperate or defect by determining which choice maximizes their *expected* share of the good for the remainder of the game. In addition, uncertainty is introduced into the relation between individual effort and group performance to model imperfect information and bounded rationality. The incorporation of diversity into the model is left until Section 4, so that its effects may be more apparent.

The type of expectations that we consider consist of two components: each individual believes (1) that future aggregate collective behavior is directly influenced by the individual's choices in inverse proportion to the group size; and (2) that the interaction is of finite duration characterized by a horizon length, *H*. Thus, individuals believe that in the long run their actions encourage similar actions on the part of others in the group. There are various mechanisms by which this might occur: imitation and establishment of conventions or norms are a few. In addition, individuals believe that the ability of their actions to encourage like actions increases with the number of individuals who have chosen to contribute to the collective good (as opposed to free riding). In some sense, this reflects a belief that contributing individuals form a core group whose reactions are much more sensitive to fluctuations in the amount of the good produced.

We show that an individually rational strategy of conditional cooperation emerges from the individuals' expectations and beliefs. Individuals cooperate if they perceive the fraction cooperating to be greater than some critical amount and defect otherwise. This strategy of conditional cooperation is reminiscent of the successful tit-for-tat strategy in 2–player prisoners' dilemma in which a player cooperates if and only if its opponent cooperated in the previous turn and defects otherwise [11].

In Section 3, we analyze the dynamics of fluctuations away from Nash equilibria using a thermodynamic-like formalism [12] for uniform groups without diversity. Besides confirming that there exists a critical group size beyond which cooperation is not sustainable [1, 9], we find additional, intriguing effects. Our results reveal several different dynamical regimes that should be observable as the size of a group changes. In one regime, although cooperation may persist for very long times even for groups exceeding a critical size, group behavior eventually decays to overall defection. In another situation a system can be stuck in a non-cooperative state even though its size is well below that guaranteeing long term cooperation. These effects are shown to depend strongly on the degree of uncertainty pervading the system, as well as on the length of the individual's horizons. To confirm our analytical predictions, we present the results of computer simulations.[1]

In Section 4 we study the effect of diversity among individuals on the phenomena described in Section 3. We find that without changing the critical sizes for cooperation, diversity acts a source of additional uncertainty and shortens the transition time to the long-term stable state of the system. In fact, an analytical form for the increase in uncertainty due to diversity is given. Again, computer simulations are invoked which confirm our predictions. Moreover, these simulations also show that when there is diversity of beliefs among subgroups merged to form a large group, cooperation can be achieved in stages.

Section 5 summarizes our results.

---

[1] The interested reader is referred Glance and Huberman [13] for additional technical detail beyond that provided in this paper.

## 2 The Dynamics of Collective Action

**The economics of free riding**

We consider the ongoing group interaction involved in the production of a non-excludable collective good which exhibits limited jointness of supply. Each individual can either contribute (cooperate) to the production of the good, or not (defect). While no individual can directly observe the effort of another, each member observes instead the collective output and can deduce overall group participation using knowledge of individual and group production functions. We also introduce an amount of uncertainty into the relation between members' efforts and group performance. There are many possible causes for this uncertainty [9]; for example, a member may try but fail to contribute due to unforeseen obstacles. Alternatively, another type of uncertainty might arise due to individuals with bounded rationality occasionally making suboptimal decisions [14, 15]. In any case, we treat here only idiosyncratic disturbances or errors, whose occurrences are purely uncorrelated.

Consequently, we assume that a group member intending to participate does so successfully with probability $p$. The individual's attempt to cooperate, instead of guaranteeing contribution to the cause, fails with probability $1-p$, with an effect equivalent to a defection. The inverse scenario holds for members intending to defect: an attempt to defect results in zero contribution with probability $q$, but with probability $1-q$ a defecting member will inadvertently (or suboptimally) contribute to the collective good. Then, as all attempts are uncorrelated, the number of successfully cooperating members, $\widehat{n}_c$, is a mixture of two binomial random variables with mean $<\widehat{n}_c> = pn_c + (1-q)(n-n_c)$, where $n_c$ is the number of members attempting to cooperate within a group of size $n$.

The limit $p$ and $q$ equal to 1 corresponds to an error-free world of complete information, while $p$ and $q$ equal to 0.5 reflect the case where the effect of an action is completely divorced from intent. Note that whenever $p$ and $q$ deviate from 1, the perceived level of cooperation will differ from the actual attempted amount of participation.

In the simplest case, collective benefits increase linearly in the contributions of the members, at a rate $b$ per cooperating member. Each contributing individual bears a personal cost, $c$. Let $k_i$ denote whether member $i$ is cooperating ($k_i$=1) or defecting ($k_i$=0). Then the utility at time $t$ for member $i$ is

$$U_i(t) = \frac{b}{n}\widehat{n}_c(t) - ck_i. \tag{2.1}$$

Using its knowledge of the functional form of the utility function[2], each individual can deduce the number of individuals effectively cooperating at some time $t$ by inverting Eq. 2.1:

$$\widehat{n}_c(t) = \frac{n}{b}(U_i(t) + ck_i). \qquad (2.2)$$

Of course, this estimation will differ from the actual number of individuals intending to cooperate in a manner described by the mixture of two binomial distributions. We also define $\widehat{f}_c(t)$, to denote the fraction, $\widehat{n}_c(t)/n$, of individuals effectively cooperating at time $t$.

When all members contribute successfully, each receives net benefits $bn/n - c = b - c$. The production of the collective good becomes an $n$-person prisoners' dilemma when

$$b > c > \frac{b}{n}. \qquad (2.3)$$

Thus, although the good of all is maximized when everyone cooperates $(b - c > 0)$, the dominant strategy in the one-shot game is to defect since additional gain of personal participation is less than the private cost $(b/n - c < 0)$.

The logic behind the decision to cooperate or not changes when the interaction is ongoing since future expected utility gains will join present ones in influencing the rational individual's decision to contribute or not to the collective good. In particular, individual expectations concerning the future evolution of the game can play a significant role in each member's decisions. The importance individuals place on the future depends on how long they expect the interaction to last. If they expect the game to end soon, then, rationally, future expected returns should be discounted heavily with respect to known immediate returns. On the other hand, if the interaction is likely to continue for a long time, then members may be wise to discount the future only slightly and make choices that maximize their returns on the long run. Notice that making present choices that depend on the future is rational only if, and to the extent that, a member believes its choices influence the decisions others make.

In the next section we elaborate on one self-consistent set of beliefs that permits individuals engaged in ongoing collective action to make the decision whether or not to contribute.

---

[2] For a justification of the form of the individual utility function in the context of either divisible goods or pure public goods see [9].

**Expectations**

The time scale of the interaction is set by the rate, $\alpha$, at which members of the group reexamine their choices. Information about the level of cooperation is deduced from individual utility accrued in the past, as per Eq. 2.2, and is thus delayed by an interval $\tau$. Along with expectations about the future, the two parameters, $\alpha$ and $\widehat{f}_c$, the fraction observed as cooperating, determine how individuals expect the level of cooperation to evolve in time.

For simplicity we assume all members of the group share a common rationality in their method of forming expectations. Specifically, all members expect the game to be of finite duration $H$, the horizon length[3]. Thus, future returns expected at a time $t'$ from the present are discounted at a rate $e^{-t'/H}$ with respect to immediate expected returns. Secondly, each member expects that their choice of action, when reflected in the net benefits received by the others, will influence future levels of cooperation. Since, however, the decision of one individual affects others' returns by an increment or decrement of only $b/n$, each member perceives its influence as decreasing with increasing group size. Furthermore, individual changes in strategy are believed to be most effective in encouraging similar behavior when levels of cooperation are high. We postulate that these two effects compound so that each member expects its decision to cooperate or defect to encourage an overall growth or decay in the level of cooperation at a rate proportional to the ratio $\widehat{f}_c/n$. Roughly, then, a member expects its cooperative (defecting) action to stimulate an additional $\widehat{f}_c/n$ members to cooperate (defect) during each subsequent time period.[4]

A mathematical formulation of the manner in which cooperation encourages cooperation and defection encourages defection with nice asymptotic properties now follows. Let $\Delta \widehat{f}_c(t + t')$ denote the expected future difference (at time $t + t'$) between the fraction of agents cooperating and the fraction of those defecting. Member $i$'s choice itself causes an instantaneous difference at $t' = 0$ of $\Delta \widehat{f}_c(t, t' = 0) = 1/n$. To reflect member $i$'s expectation that during subsequent time steps (of average duration $1/\alpha$), its action will encourage $\widehat{f}_c/n$ additional members per time step to behave likewise, we stipulate that the difference $\Delta \widehat{f}_c(t + t')$ approach 1 asymptotically from its initial value $1/n$ at some given rate. For the purposes of simplicity we will set this rate equal to $e^{-\alpha \widehat{f}_c(t-\tau)t'/n}$,

---

[3] The concept of a horizon is formally related to a discount $\delta$, which reflects the perceived probability that the game will continue through the next time step. The two are connected through the relation $\sum_{i=0}^{\infty} \delta^i\, () \to \int_0^{\infty} dt'\, e^{-t'/H}\, ()$ which implies $H = \frac{1}{1-\delta}$.

[4] Assigning a functional form to individual expectations, although somewhat arbitrary, is necessary to study dynamics. Variations on the same theme, however, can be seen to yield similar dynamics, provided that the rate at which similar behavior is expected to encourage similar behavior rises monotonically in the ratio $\widehat{f}_c/n$.

which corresponds to the following deviation:

$$\Delta \widehat{f}_c(t+t') = 1 - (1-1/n)\exp\left(-\frac{\alpha \widehat{f}_c(t-\tau)}{n}t'\right). \tag{2.4}$$

We should point out that variations on the precise functional form of the expected deviation $\Delta \widehat{f}_c(t+t')$ would simply cause to the deviation to grow faster or slower with increasing $t'$, without yielding significant qualitative changes in the types of dynamical behavior characterizing the interaction.

In summary, member $i$ reevaluates its decision whether or not to contribute to the production of the good at an average rate $\alpha$, using its knowledge of $\widehat{f}_c(t-\tau)$ and following its expectations about the future. Using its prediction of how it expects $f_c$ to evolve in relation to its choice and discounting the future appropriately, member $i$ then makes its decision on whether to cooperate or defect.

Putting it all together, member $i$ perceives the advantage of cooperating over defecting at time $t$ to be the net benefit

$$\Delta \widehat{B}_i(t) = H(b-c) - \frac{Hb(n-1)}{n+H\alpha \widehat{f}_c(t-\tau)}. \tag{2.5}$$

Member $i$ cooperates when $\Delta \widehat{B}_i(t) > 0$, defects when $\Delta \widehat{B}_i(t) < 0$, and chooses at random between defection and cooperation when $\Delta \widehat{B}_i(t) = 0$, basing its decision on the fraction of the group it perceives to have cooperated at a time $\tau$ in the past, $\widehat{f}_c(t-\tau)$. These criteria reduce to the following condition for cooperation at time $t$:

$$f_{crit} \equiv \frac{1}{H\alpha}\left(\frac{nc-b}{b-c}\right) < \widehat{f}_c(t-\tau). \tag{2.6}$$

Since $\widehat{f}_c(t-\tau)$ is a mixture of two binomially distributed variables, Eq. 2.6 provides a full prescription of the stochastic evolution for the interaction. In particular, member $i$ cooperates with probability $P_c(f_c(t-\tau))$ that it perceives cooperation as maximizing its expected future accumulated utility, given the actual attempted level of cooperation $f_c(t-\tau)$.

Thus, the individuals engage in conditional cooperation, cooperating when the fraction perceived as cooperating is greater than the critical amount, $f_{crit}$ and defecting when the perceived fraction cooperating is less than $f_{crit}$. Because of the nature of the individuals' expectations, this behavior, reminiscent of a generalized tit-for-tat, emerges from individually rational choices. In this way, conditional cooperation is a rational strategy. Furthermore, conditional cooperation is a more general type of strategy than might be expected from the explicit specification of the individuals' beliefs. As a result, it holds for a wider range of models than considered here, once one allows for variation in the functional form of $f_{crit}$.

**Group behavior**

A deterministic continuous version of the stochastic discrete interaction specified above can be obtained assuming: (1) the size, $n$, of the group is large; and (2) the average value of a function of some variable is well approximated by the value of the function at the average of that variable. The model developed below will be useful in discovering the Nash equilibria and determining their stability characteristics.

From this point on, we specialize to the symmetric case $p = q$, in which an individual is equally likely to effectively defect when intending to cooperate as to cooperate when intending to defect. (Later we will comment on the effect of this restriction on the generalizability of our results.) By the Central Limit Theorem, for large $n$, the random variable $\widehat{f}_c$ tends in law to a Gaussian distribution with mean $\langle \widehat{f}_c \rangle = pf_c + (1-p)(1-f_c)$ and variance $\sigma^2 = p(1-p)/n$. Under our assumptions, the mean probability that $\widehat{f}_c > f_{crit}$ thus becomes

$$\rho_c(f_c) = \frac{1}{2}\left\{1 + \mathrm{erf}\left[\frac{\langle \widehat{f}_c \rangle - f_{crit}}{\sqrt{2}\sigma}\right]\right\}. \tag{2.7}$$

The evolution of the number of agents cooperating in time is then described by the dynamical equation [16]

$$\frac{df_c}{dt} = -\alpha[f_c(t) - \rho_c(f_c(t-\tau))], \tag{2.8}$$

where $\alpha$ is the reevaluation rate and $\tau$ is the delay parameter, as defined earlier.

The equilibrium points $f_c^0$ of the interaction described by the above equation are obtained by setting the right hand side to zero[5]. In this case they are given by the

---
[5] Furthermore, linear stability analysis of Eq. 2.8 shows that the stability of the equilibrium points is independent of the value of the delay $\tau$, and of the reevaluation rate $\alpha$. Thus, the asymptotic behavior of the group interaction does not depend on the delay. Moreover, the equilibrium points belong to one of two types: stable fixed point attractors or unstable fixed point repellors, due to the linearity of the condition for cooperation (Eq. 2.6).

solutions to

$$\rho\left(f_c^0\right) = f_c^0. \tag{2.9}$$

Solving the above equation yields the critical sizes beyond which cooperation can no longer be sustained. In the case of perfect certainty ($p=q=1$), these critical sizes can be expressed in simple analytical form. Thus, a group will no longer sustain global cooperation if it exceeds a value $n^*$ given by

$$n^* = H\alpha\left(\frac{b}{c} - 1\right) + \frac{b}{c}. \tag{2.10}$$

Similarly, cooperation is the only possible global outcome if the group size falls below a second critical size $n_{min}$:

$$n_{min} = \frac{b}{2c} + \frac{1}{2c}\sqrt{b^2 + 4H\alpha c(b-c)}. \tag{2.11}$$

Notice that these two critical sizes are not equal; in other words, there is a range of sizes between $n_{min}$ and $n^*$ for which either cooperation or defection is a possible outcome, depending on the initial conditions.

An estimate of the possible sizes can be obtained, for example, if one assumes a horizon $H=50$ (which corresponds to a termination probability $\delta=0.98$), $\alpha=1$, $b=2.5$ and $c=1$. In this case one obtains $n^* = 77$ and $n_{min} = 10$, a significant group size. Observe that an increase in the horizon length would lead to corresponding increases in the critical sizes.

# 3 The Critical Mass

**The $\Omega$ function formalism**

The model studied in the previous section dealt with the average properties of a collection of agents having to choose between cooperation and defection. Since the asymptotic behavior generated by the dynamics was in the form of Nash equilibria or fixed points, it is of interest to ask about the evolution of fluctuations away from equilibrium state in the presence of uncertainty. These fluctuations are important for two reasons: (1) the time necessary for the system to relax back to equilibrium after small departures in the number of agents cooperating or defecting might be long compared to the time-scale of the collective task to be performed, or the measuring time of an outside observer; (2) large enough fluctuations in the number of defecting or collaborating agents can shift the state of the system from cooperating to defecting and *vice-versa*. If that is the case, it becomes important to know how probable these large fluctuations are, and how they evolve in time.

In what follows we will use a formalism introduced by Ceccatto and Huberman [12] that is well suited for studying fluctuations away from the equilibrium behavior of the system. This formalism relies on the existence of an optimality function, $\Omega$, that can be constructed from knowledge of the density dependent utilities. This function has the important property that its local minima give the Nash equilibria of the system as the most probable configurations of the system. Depending on the complexity of the function, several Nash equilibria can exist, with the overall global minimum producing the optimal state of the system.

Specifically, the equilibrium probability distribution $P_e(f_c)$ is given by

$$P_e = C \exp\left[-n\Omega(f_c)\right], \tag{3.1}$$

where the optimality function $\Omega$ for our model of ongoing collective action is given by

$$\Omega(f_c) = \int_0^{f_c} df'_c \left[f'_c - \rho_c(f'_c)\right] \tag{3.2}$$

in terms of the mean probability $\rho_c(f_c)$ that cooperation is preferred. Thus, the optimal configuration corresponds to the value of $f_c$ at which $\Omega$ reaches its global minimum.

Within this formalism it is easy to study the dynamics of fluctuations away from these minima. First, consider the case where there is a single Nash equilibrium (which can be either cooperative or defecting). As was shown, fluctuations away from this state relax back exponentially fast to the equilibrium point, with a characteristic time of the order of $1/\alpha$, which is the average evaluation time for the individuals. Second, and more interestingly, is the situation when there are multiple Nash equilibria, with the global minimum of the $\Omega$ function denoting the optimal state of the system. This is illustrated schematically in Fig. 1.

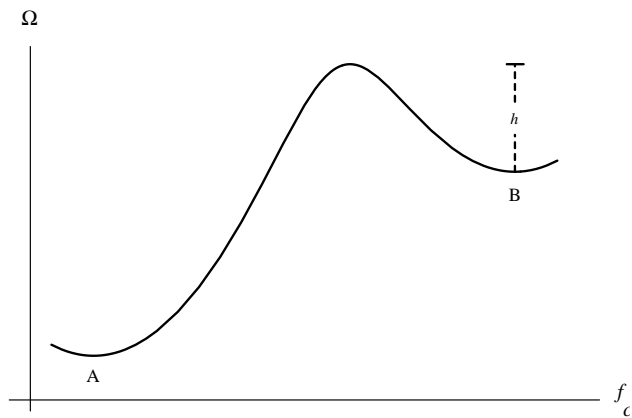

**Fig. 1.** Schematic sketch of the optimality function $\Omega$ vs. $f_c$, the fraction of agents cooperation. The global minimum is at *A*, local minimum at *B*. *h* is the barrier height separating state *B* from *A*.

If the system is initially in a Nash equilibrium which corresponds to the global minimum (*e.g.*, state *A*), fluctuations away from this state will relax back exponentially fast to that state. But if the system is initially trapped in a metastable state (state *B*), *i.e.*, a minimum which is not the global one, the dynamics away from this state is both more complicated and interesting. As was shown by Ceccatto and Huberman, whereas for short times fluctuations away from the local minimum relax back to it, for longer times a giant fluctuation can take place, whereby a large fraction of the agents switching strategies can push the system over the barrier maximum. Once this critical mass is reached, the remaining agents rapidly switch into the new strategy that corresponds to the optimal Nash equilibrium and the system slides into the optimal state.

The time scales over which this whole process takes place is also of interest, for the time to nucleate a giant fluctuation is exponential in the number of agents. However, when such transitions take place, they do so very rapidly — the total time it takes for all agents to do the crossing is logarithmic in the number of agents. Since the logarithm of a large number is very small when compared to an exponential of the same number, the theory predicts that nothing much happens for long times, but when it does, it happens very fast.

The process of escaping from the metastable state depends on the amount of imperfect knowledge that individuals have about the state of the system, in other words, on what other agents are doing. In the absence of imperfect knowledge the system would always stay in the local minimum downhill from the initial conditions, since small excursions away from it by a few agents would reduce their utility. It is only in the case of imperfect knowledge that many individuals can change their behavior. This is because, in evaluating the number of members cooperating, imperfect knowledge amounts to occasional large errors in the individual's estimation of the actual number cooperating.

Results of Monte Carlo simulations, which were conducted in asynchronous fashion, confirm these theoretical predictions. Each individual decides to cooperate or defect based on the criterion given in Eq. 2.6. Uncertainty enters since these decisions are based on perceived levels of cooperation which differ from the actual attempted amount of cooperation in a way distributed as a mixture of binomials.

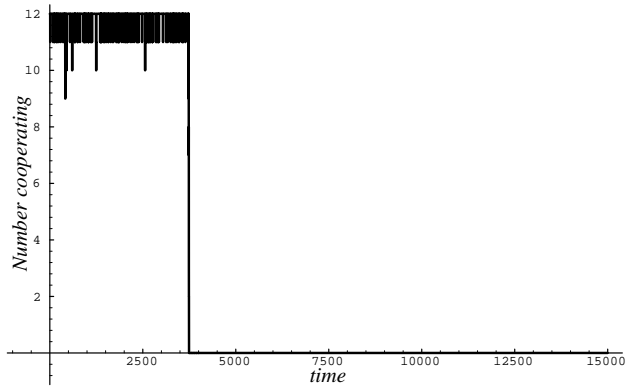

**Fig. 2.** At $t=0$, two cooperating groups of size $n=6$ merge to form a larger, cooperating group of size $n=12$. All agents have horizon length $H=9.5$, with $p=0.93$, $b=2.5$, $c=1$, $\alpha=1$, and $\tau=1$. For these parameters, cooperation is the optimal state for a group of size $n=6$, but for the combined group of size $n=12$, cooperation is metastable. Indeed, as the figure shows, metastable cooperation persists for almost 4,000 time steps in this example. The average crossover time is about 5,000 time steps, ranging from less than 1,000 to over 10,000 time steps. Uncertainty ($p$ less than one) ensures that eventually a large fluctuation in the perceived number of agents cooperating eventually takes the group over into a state of mutual defection, which is optimal.

For example, consider two small cooperating groups of size $n=6$, with horizon length $H=9.5$, for which the optimal state (*i.e.*, the global minimum of the optimality function) is cooperation, that merge at $t=0$ to form a larger, cooperating group of size $n=12$. For the larger group, cooperation is now a metastable state: no one individual will find it to its benefit to defect and the metastable cooperative state can be maintained for very long times, especially if *p* is close to 1. As shown in Fig. 2, in this one case mutual cooperation lasts for about 4000 time steps, until a sudden transition (of duration proportional to the logarithm of the size of the group) to mutual defection occurs, from which the system will almost never recover (the time scale of recovery is many orders of magnitude larger than the crossover time).

Determining the average time that it takes for the group to crossover to the global minimum is a calculation analogous to particle decay in a bistable potential and has been performed many times [17]. The time, *t*, that it takes for a group of size *n* to cross over from a metastable Nash equilibrium to the optimal one is given by

$$t = \text{constant } e^{nh/\sigma}, \qquad (3.3)$$

with *h* the height of the barrier as shown in Fig. 1 and $\sigma$ a measure of the imperfectness of the individuals' knowledge. We should point out, however, that in our model the barrier height itself also depends on *n*, *H*, and *p*, making simple analytical estimates of the crossover time considerably more difficult.

Further simulations of the example given above show that the average crossover time in that case is about 5,000 time steps, although it can range from less than 1,000 to over 10,000 time steps. The exponential dependence of the crossover time on the amount of uncertainty can be seen by running the same system, but with different amounts of error. In the example above, *p* equals 0.93. However, if the amount of error increases so that *p* now equals 0.91, say (thus reducing the height of barrier between cooperation and defection by 21%), the crossover to defection typically occurs within hundreds of time steps, instead of thousands.

**Critical sizes for cooperation**

The optimality function reveals much of what we wish to know concerning the dynamics of a system engaged in a collective action problem: it gives the possible Nash equilibria and predicts the long-term stable state, *i.e.*, the state that corresponds to the global minimum. It also shows that as the size of the group changes, the relative depths of $\Omega$'s minima change, leading to new optimal states.

Eq. 2.6 implies that $f_{crit}$ increases with increasing *n*. As a result, the value of $f_{crit}$ passes from $f_{crit} < 0.5$ to $f_{crit} > 0.5$ as *n* increases. This indicates a transition in the dynamical nature of the interaction: as *n* increases, the interaction switches from having an optimal state of mutual cooperation to having an optimal state of mutual defection. These transition points were derived exactly for *p*=1 and numerically for *p*<1.

As $p$ decreases from 1, numerical analysis of $\Omega(f_c)$ shows that the shape of the optimality function is roughly preserved, but that the barrier height decreases with decreasing $p$. The peak drifts away from $f_c = f_{crit}$ (except in the special case $f_{crit}$=0.5), while the minima may move in from $f_c = 0$ and $f_c = 1$. Eventually, as $p$ is decreased further, the barrier height goes to zero at some $p_{crit}$, and only one minimum remains.

Until $p$ reaches the critical value $p_{crit}$ at which the barrier height goes to zero for all $n$, the nature of the system's equilibrium points remain similar to the $p$=1 case. Specifically, for $p < p_{crit}$, three critical values can be obtained: (1) $n_{min}(p)$, the minimum group size below which cooperation is the only fixed point; (2) $\tilde{n}^*(p)$, the critical size below which cooperation is the optimal state; and (3) $n^*(p)$, an upper bound above which cooperation is not sustainable. The values $n_{min}(p)$, $\tilde{n}^*(p)$, and $n^*(p)$ can be determined numerically, demonstrating the emergence of four levels in group size corresponding to very different resolutions of the collective action problem:

| | |
|---|---|
| $n \leq n_{min}(p)$ | one equilibrium point – mostly cooperative; |
| $n_{min}(p) < n \leq \tilde{n}^*(p)$ | mostly cooperative optimal state, mostly defecting metastable state; |
| $\tilde{n}^*(p) < n < n^*(p)$ | mostly defecting optimal state, mostly cooperative metastable state; |
| $n \geq n^*(p)$ | one equilibrium point – mostly defecting. |

Analysis of the optimality function yields the values $n_{min}(p)$, $\tilde{n}^*(p)$, and $n^*(p)$, along with $p_{crit}$. These values were found for the case $H$=50 (which corresponds to discount rate $\delta = 1 - \frac{1}{H} = 0.98$), $\alpha$=1, $\tau$=1, benefit to group of individual cooperation $b$=2.5, and personal cost of cooperation $c$=1. The resulting phase diagram delineating the regions of different resolutions to the conflict is shown in Fig. 3. At $p$=1, $n^* = 77, \tilde{n}^* = 40$, and $n_{min} = 10$. When $p > p_{crit} = 0.59$, the uncertainty is high enough that all structure in the optimality function is washed out and only one equilibrium point exists for group of all sizes. For groups of size less than 40 operating within such a high level of uncertainty, the interaction evolves to a mixed equilibrium that is more cooperative than defective. The balance reverses itself when the group size exceeds 40.

The diagram in Fig. 3 shows that for the symmetric case $p = q$, $n^*(p)$ is a decreasing function of $p$ while $n_{min}(p)$ is an increasing function of $p$ and $\tilde{n}^*(p)$ is a constant. In general $(q \neq p)$, $n^*(p)$ remains a decreasing function; however, the functional forms of $\tilde{n}^*(p)$ and $n_{min}(p)$ depend on $q$. For example, $q$=0 yields decreasing functions $\tilde{n}^*(p)$ and $n_{min}(p)$.

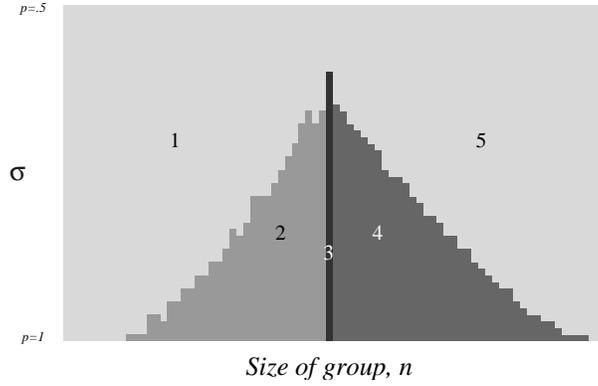

**Fig. 3.** Diagram delineates regions corresponding to different resolutions of collective action problem for parameter values $H=50$ and $\alpha=1$. The amount of error, $\sigma$, increases vertically, and the size of the group increases horizontally. In region 1, $n \leq n_{min}(p)$, there is one equilibrium point – mostly cooperative; in region 2, $n_{min}(p) < n < \tilde{n}^*(p)$, mostly cooperative is the optimal state, while mostly defecting is a metastable state; in region 3, $n = \tilde{n}^*(p)$, the system is bistable with mostly cooperative and mostly defecting both optimal states; in region 4, $\tilde{n}^*(p) < n < n^*(p)$, mostly defecting is the optimal state, while mostly cooperative is a metastable state; and in region 5, $n^*(p) \leq n$, there is again only one equilibrium point – mostly defecting. There is no sharp boundary between regions 1 and 5 for high levels of uncertainty. Note that region 3 actually has zero width.

# 4 Diversity and Cooperation

In the first pass at studying collective action problems, we treated all individuals as identical. We now drop that assumption and consider how diversity enters and how behavior changes as a result. As we discuss below, two qualitatively different forms of diversity must be studied. The first type reflects diversity in groups of agents whose differences in beliefs can be captured by a simple spread about some common belief. Thus, the individuals on the whole are similar, but have differences that capture variability in preferences and other additional factors. For example, each group member might be said to have horizon length $H = 10 \pm 2$, instead of $H=10$ exactly.

On the other hand, the second type of diversity represents differences within a group that cannot be accounted for by a simple variance about an average value. Instead the group acts as the union of several subgroups each characterized by its own set of beliefs. In this case, one subgroup might have horizon length $H=2$, another $H=4$, and yet another $H=6$.

A general way of incorporating diversity is to allow the critical fraction, $f_{crit}$, in the criterion for cooperation (Eq. 2.6) to vary from individual to individual. Thus, each member $i$ has a bias $b_i$, and decides whether or not cooperate based on the revised condition

$$f_{crit} + b_i < \widehat{f}_c(t - \tau). \tag{4.1}$$

The notation we use to denote the biased value of the critical fraction will be $f_{crit}^{b_i} \equiv f_{crit} + b_i$.

**Diversity as a form of uncertainty**

We examine first the case in which the group's diversity can be modeled by a Gaussian distribution. If we take the biases $\{b_i\}$ to be distributed normally with mean zero, then the critical size beyond which cooperation cannot be maintained remains the same as without diversity. Distributions with non-zero mean would alter the nature of the equilibrium points; so zero mean distributions offer the best means to isolate the effect of adding diversity.

In addition, the distribution of biases $\{b_i\}$ can be used to represent a diversity in the individuals' horizons or in their benefit-cost ratios $b/c$ as long as $\sigma \ll f_{crit.}$. Thus, describing diversity by adding a bias to the criterion for cooperation of Eq. 2.6 is a more general description of diversity than might appear at first glance.

In order to understand the qualitative effect of introducing diversity in this manner, we write the mean probability $\rho_c^i(f_c)$ that member $i$ will choose to cooperate as

$$\rho_c^i(f_c) = \frac{1}{2}\left\{1 + erf\left[\frac{\langle \widehat{f}_c \rangle - (f_{crit} + b_i)}{\sqrt{2}\sigma}\right]\right\}, \qquad (4.2)$$

where $\langle \widehat{f}_c \rangle = pf_c + (1-p)(1-f_c)$ and $\sigma = \sqrt{p(1-p)/n}$. The dynamical equation describing the evolution of the system then becomes

$$\frac{df_c}{dt} = -\alpha\left[f_c(t) - \frac{1}{n}\sum_{i=1}^n \rho_c^i(f_c(t-\tau))\right]. \qquad (4.3)$$

Letting

$$\tilde{\rho}_c(f_c) = \frac{1}{n}\sum_{i=1}^n \rho_c^i(f_c(t-\tau)). \qquad (4.4)$$

and linearizing about $\langle \widehat{f}_c \rangle = f_{crit}$, we obtain

$$\tilde{\rho}_c(f_c) = \frac{1}{2}\left\{1 + erf\left[\frac{\langle \widehat{f}_c \rangle - f_{crit}}{\sqrt{2}\tilde{\sigma}}\right]\right\}, \qquad (4.5)$$

with the renormalized value of uncertainty $\tilde{\sigma}$ given by

$$\tilde{\sigma} \to \sqrt{\sigma^2 + \sigma'^2}. \qquad (4.6)$$

This approximation assumes that the $\{b_i\}$ are distributed normally with mean zero and standard deviation $\sigma'$.

Thus, taking this first form of diversity into account simply renormalizes the amount of noise in the system as parametrized by $\sigma$ in the denominator of the error function. Note, however, that imperfect information also enters in the calculation of the expected value of $f_c$: $\langle \widehat{f}_c \rangle = pf_c + (1-p)(1-f_c)$. Consequently, adding diversity is not identical to decreasing $p$ away from 1.

From this analysis, it appears that diversity among the agents will shorten the lifetime of the metastable states described in the previous section, while the transitions remain abrupt. Computer experiments verify this prediction, with a sample simulation given in Fig. 4. Diversity is modeled as a spread about $f_{crit}$ among the agents; thus, agents differ from one another in their likelihood of cooperating *vs.* defecting. In this example, the individual biases to $f_{crit}$ are distributed normally with standard deviation set equal to the amount due to imperfect information $\left(\sigma' \equiv \sigma = \sqrt{p(1-p)/n}\right)$. As a result of the diversity among the individuals in the group, the average crossover time becomes about 1,300 time steps, ranging from fewer than 100 time steps to over 2,000. Without diversity, the average crossover time is about 5,000 time steps (see, for example, Fig. 2 — note in particular the difference in time scales), ranging from less than 1,000 time steps to over 10,000.

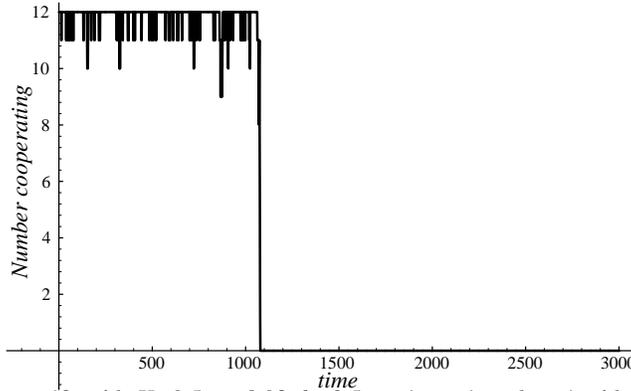

**Fig. 4.** Group of size $n=12$, with $H=9.5$, $p=0.93$, $b=2.5$, $c=1$, $\alpha=1$, and $\tau=1$ with diversity. Diversity is modeled by adding individual biases to the value of $f_{crit} = 0.6667$. These biases are distributed normally with mean zero and standard deviation $\sigma' \equiv \sigma = \sqrt{p(1-p)/n}$ (the "noise" due to diversity is set equal to the noise due to uncertainty). The moderate amount of diversity is responsible for a transition rate to defection about four times as fast as that for a uniform group (see, for example, Fig. 2 — note the difference in time scales).

## Stages to cooperation

We next consider the second form of diversity, representing large differences in beliefs between various subgroups within a group. This type of diversity follows a multimodal distribution and does not lend itself to the renormalization of uncertainty performed above. Instead, the summed error function in Eq. 4.3 exhibits a sequence of steps. For example, if in a group of size 12 there is one subgroup biased towards cooperation with $f_{crit}^{b_1} = 0.1333$, another with $f_{crit}^{b_2} = 0.4667$ and yet another biased towards defection with $f_{crit}^{b_3} = 0.8$, then there are three steps in the summed error function of Eq. 4.4. Each step represents the likelihood that an individual from each respective subgroup will find it worthwhile to cooperate. There can now be multiple local minima in which some of the subgroups sustain cooperation while the others do not. As always, the global minimum is the long-term preferred state of the system.

In the example given in the preceding paragraph, overall cooperation can be achieved in step-wise fashion. If the group begins in an uncooperative state, the subgroup with $f^{b_1}_{crit} = 0.1333$ is the first to make the transition to cooperation, followed by the subgroup with $f^{b_2}_{crit} = 0.4667$ and finally by the subgroup with $f^{b_3}_{crit} = 0.8$. These steps can be seen in Fig. 5.

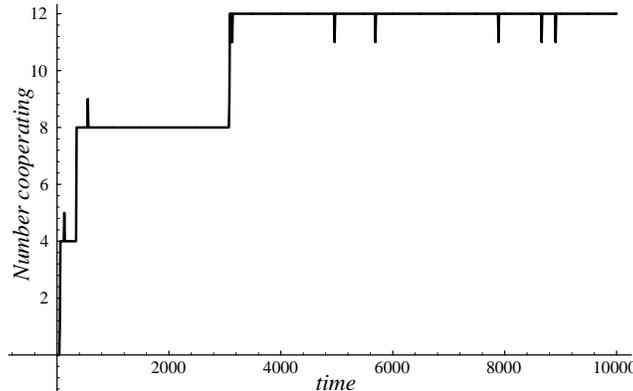

**Fig. 5.** *Stages towards cooperation.* A group of size 12 consists of three subgroups. The first is greatly biased towards cooperation with $f^{b_1}_{crit} = 0.1333$. The second subgroup has $f^{b_2}_{crit} = 0.4667$, while the third is biased towards defection with $f^{b_3}_{crit} = 0.8$. The figure shows the stepwise transition to mutual cooperation: each subgroup makes the transition to cooperation in turn. Parameter values are $p=0.98$, $b=2.5$, $c=1$, $n=12$, and $H=12$.

Compare this example to the uniform case in which all individuals have $f_{crit} = 0.4667$ (the average critical fraction for the diverse group above). This system is now almost bistable, with both cooperation and defection about equally likely to be the long-term stable state, although cooperation is more likely. In this scenario, then, diversity greatly increases the likelihood of attaining a cooperative state. Other scenarios can be envisioned, however, which accomplish the reverse, making defection more likely with diversity than without.

Another interesting consequence of the effect of multimodal diversity is the following. Imagine combining two groups, each composed of 6 individuals. The first group has beliefs such that all members have $f^b_{crit} = 0.3667$. Thus, the long-term stable state for this group is one of cooperation. The second group, on the other hand, is unbiased with $f_{crit} = 0.6667$, and thus tends towards mutual defection. Putting these groups together, then, we have an initial state with 6 of the 12 individuals cooperating. This turns out to be a metastable state of the combined system. Eventually, however, the group undergoes a transition to either complete cooperation or complete defection. Both are equally likely — the two possibilities are illustrated in Fig. 6. Thus, either the defecting agents are persuaded to cooperate (although they generally wouldn't when in a group to themselves even though the size of this subgroup is smaller); or the originally cooperating agents lose their incentive to do so and defect.

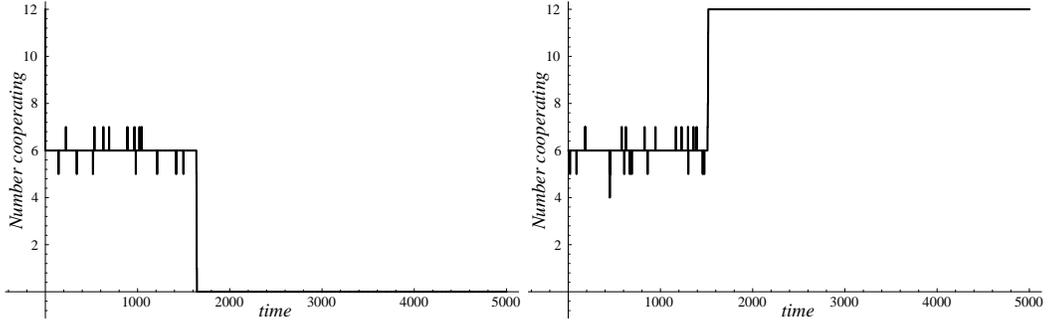

**Fig. 6.** Two groups of size 6 are joined together, one with a bias towards cooperation $\left(f^{b_1}_{crit} = 0.3667\right)$, the second with a tendency to defect. ($f_{crit} = 0.6333$). Thus, the long-term state for the first group on its own is cooperation, while the long-term state for the second is defection. Combining the two groups with the six agents biases towards cooperation initially cooperating and the second six defecting, the figures show that overall behavior can go either way. The initial state is itself metastable, but fluctuations insure that eventually the group will go over to either mutual cooperation or mutual defection, with both possibilities equally likely. Either the defecting agents are swept over into cooperation or the cooperative agents collapse into defection. Parameter values are $p$=0.99, $b$=2.5, $c$=1, $n$=12, $H$=12.

This example invites comparison to the situation in which all members have $f_{crit} = 0.5666$. Such a group would be nearly bistable, although with a long-term tendency towards defection. A group with initially half its members cooperating would most of time evolve rapidly to a state of mutual defection. With diversity as above, on the other hand, the two stable states of defection and cooperation become separated by a metastable state in which half of the group cooperates, and the group can now remain stuck in a situation in which half of the group free rides on the other half. For this particular choice of diversity, transitions to either overall cooperation or defection are equally likely. Thus, the diversity in the group has improved the group's chances of mutual cooperation over those of a uniform group with the same average $f_{crit}$.

# 5 Discussion

In this paper we presented and studied a model of ongoing collective action among intentional agents which make choices that depend not only to the past but also on their expectations. This model maps collective action problems onto an iterated $n$-person prisoners' dilemma in an uncertain world. Individuals interact dynamically, making decisions based both on individual preferences and on their expectations as to how their choices will affect other agents in the future. This is a significant departure from previous studies of collective action problems in which the finiteness of the interaction is the only shadow cast by the future.

Using a dynamical formulation of these interactions we derived the magnitudes of group sizes that are capable of supporting cooperation and established the existence of regimes with multiple Nash equilibria for groups of identical agents. Extensive computer experiments also allowed us to confirm the analytically derived behavioral diagrams whose distinct regions correspond to different resolutions of the collective action problem.

We also studied stochastic fluctuations in the number of individuals cooperating or defecting using a thermodynamic-like formalism. This allowed us to discover the existence of very long transient states in which collaboration can persist in groups whose sizes are too big to support it indefinitely. Moreover, the onset of overall cooperation takes place in a sudden and unexpected way. Likewise, defection can appear out of nowhere in very large, previously cooperating, groups. These outbreaks, which were also confirmed by computer experiments, mark the end of long transient states in which defection or cooperation persists in groups that cannot sustain it indefinitely. Since these discontinuities are not overly sensitive to the analytical form of the expectations, we expect that they might be observed in more complex social settings.

Relaxing the assumption that the individuals were identical in their beliefs and preferences, we then elucidated the effect of diversity on the dynamics of collective action. One type of diversity, represented as a spread about some common average belief among individuals, was shown to act as a source of extra uncertainty, thus shortening the time to an outbreak without affecting its abrupt nature. A second form of diversity, which occurs when a group consists of several subgroups each with its own distinct beliefs was also studied. In this case, we showed that cooperation (or defection) can be achieved by the group in demarcated stages. In general, an additional metastable state appears for each new subdivision within the group.

The implications of this work for the study of cooperation in organizations can be briefly stated. In order to achieve spontaneous cooperation on a global scale, an organization should be structured into small subunits made up of individuals with a well-designed diversity of beliefs. The small size of the units allows for the emergence of sustained cooperation; diversity hastens its appearance. To the extent that these two elements, smallness and diversity, can be preserved in a hierarchical structure, we anticipate that sustained cooperation can be achieved in more complex systems as well.